\documentclass[aps,amssymb,floatfix,prd,amsmath,preprintnumbers]{revtex4}
\usepackage{graphicx}  
\usepackage{float}
\usepackage{dcolumn}   
\usepackage{bm}
\usepackage{color}
\begin{document}

\title{\bf  Phantom cosmology in an Extended Theory of Gravity}

\author{S. K. Tripathy\footnote{Department of Physics, Indira Gandhi Institute of Technology, Sarang, Dhenkanal, Odisha-759146, India, E-mail:tripathy\_ sunil@rediffmail.com} and B. Mishra \footnote{Department of Mathematics, Birla Institute of Technology and Science-Pilani, Hyderabad Campus, Hyderabad-500078, India, E-mail:bivudutta@yahoo.com }
}\affiliation{ }

\begin{abstract}
Some phantom cosmological models without big rip singularity have been constructed in a simple extended theory of gravity. In the geometrical part of the action, a minimally coupled linear function of the Ricci Scalar and the trace of the energy momentum tensor  have been considered in place of the Ricci scalar. Four Little Rip and Pseudo Rip models have been investigated where the equation of state parameter evolves asymptotically and sufficiently rapidly to $-1$. The effect of the coupling constant of the extended gravity theory on the dynamics has been discussed. Possible  wormhole solutions for the phantom models are obtained. The possibility of Big Trip in wormholes are discussed for the models.
\end{abstract}
\maketitle
\section{Introduction} 
Since the announcement of a possible late time cosmic speed up phenomenon by  Supernova project group and high- z Supernova group two decades ago \cite{Riess98, Perlmutter99}, a lot of cosmological observations from  large scale structure \cite {Tegmark2004, Abaz2004, Pope2004}, Baryon Acoustic Oscillation (BAO)  \cite{Percival10, Parkinson12}, Cosmic Microwave Background (CMB) anisotropy \cite{Sperg2003, Hinshaw13} and weak lensing \cite{Jain2003} have come up to establish the fact. The late time phenomena posed a challenge to cosmologists. In the framework of General Relativity (GR), such a phenomenon is attributed to a mysterious fluid with negative pressure, called Dark Energy (DE) (one may refer to \cite{Bamba2012} for a nice review).  GR with its simple structure $G_{\mu\nu}= \kappa T_{\mu\nu}$, has a great  success in explaining many complex issues in astrophysics and cosmology for hundred years. However it fails to explain the late time cosmic dynamics. This failure has led to the concept of modification of GR. In the context of modification of GR there can be two possibilities: either to modify the matter side by considering some additional dynamic exotic degrees of freedom or to modify the geometrical part so that the extra terms in the modified theory will provide an anti gravity effect leading to acceleration. As a simple modification, cosmological constant as a source of dark energy is a good choice and is quite compatible with recent observations. However, different DE models with exotic matter fields  such as quintessence, tachyons, phantom fields, Ricci dark energy, ghost dark energy have been proposed with some degree of success. Scalar field models proposed as a solution to the late time cosmic speed up issue are usually crippled with the inclusion of ghost fields with unusual negative kinetic energy terms at least around flat, cosmological and spherically symmetric backgrounds \citep{Bul72, Koyama07, Sbisa15, Gumru16}. Further  the intriguing and bizarre fact concerning the DE is that, it violates strong energy condition and can cluster at large scales. On the other hand, geometrically modified gravity theories have gained a lot of research attention because of non involvement of any DE candidates including ghost fields in the field equations. In these modified theories, the Ricci scalar $R$ in the gravitational action is replaced by a more general function of $R$ or by a  matter-geometry coupled functional. Out of several modified theories proposed in recent times $f(R)$ theory \cite{Caroll2004, Nojiri2007, Bert2007}, $f(G)$ gravity \cite{Nojiri2005,Li2007}, $f(\mathcal{T})$ theory \cite{Linder2010, Myrza2011} and $f(R,T)$ theory \cite{Harko2011} have gained much attention. Recently, extended gravity theory has attracted a lot of research interest because of it simple structure and ability  to reduce to GR under suitable choices of the coupling parameters \cite{Mishra18a, Mishra18b}. The $f(R,T)$ gravity theory as proposed by Harko et al. \cite{Harko2011} has been studied widely in recent times to address many issues in cosmology and astrophysics \cite{Alves16, Zubair15, Alha16, Mishra16, Yousaf16, Singh18, Tretyakov18, Velten17, Abbas17, Wu18, Baffoul17, Carvalho17, Mishra18c, Baffou19, tripathy2019}.

In the framework of GR, dark energy corresponds to an exotic fluid with negative pressure and is described by the EoS parameter $\omega=\frac{p}{\rho}$, where $p$ is the pressure of the fluid and $\rho$ is the energy density. DE models with a cosmological constant ($\Lambda$CDM model) predicts the EoS parameter as $\omega=-1$, whereas quintessence models predict $\omega > -1$. However, recent observational data favour phantom models with $\omega <-1$ over quintessence models \cite{Tripathi2017}. The 9 year WMAP survey suggests the EoS parameter to be $\omega=-1.073^{+0.090}_{-0.089}$ from CMB measurements and $\omega=-1.084\pm 0.063$ in combinations with Supernova data\cite{Hinshaw13}. Amanullah et al. of Supernova cosmology project have found that $\omega=-1.035^{+0.055}_{-0.059}$ \cite{Amanullah2010}. Kumar and Xu from a combined  analysis of the data sets of SNLS3, BAO, Planck, WMAP9 and WiggleZ constrained the EoS parameter as $\omega=-1.06^{+0.11}_{-0.13}$ \cite{Kumar2014}. Moreover the recent Planck 2018 results constrained $\omega=-1.03\pm 0.03$ \cite{Planck2018}. In view of these constraints on the EoS parameter from observations, one can not rule out the possibility of a phantom phase in the universe. On the other hand, in phantom models, energy conditions are violated and the universe evolve to a finite time future singularity. According to the classification of Nojiri et al. \cite{Nojiri2005a}, four different possibilities of singularity may occur : (i) Big Rip singularity ( type-I singularity) where the scale factor and density becomes infinite in finite time \cite{Caldwell2003}, (ii) Sudden singularity (type-II singularity) where the pressure becomes infinite while keeping the scale factor and density finite \cite{Barrow2004}, (iii)  Type-III singularity, where the pressure and density both become infinite but the scale factor remaining finite and (iv) Type-IV singularity where the higher derivatives of the Hubble parameter diverge. Inconsistencies may occur due to such finite time future singularity and to avoid such inconsistencies many scenarios have been proposed which include quantum effects to delay the singularity, modification of gravity, coupling dark energy with dark matter in a special manner or the use of specific equations of state \cite{Nojiri2011, Framp2012}. Some other models have been proposed where the dark energy density increases with time, the EoS parameter evolves asymptotically from $\omega<-1$ to $-1$ rapidly and there is effectively no finite time future singularity \cite{Framp2011, Framp2012, Asta2012}. Such models include the Little Rip (LR) and Pseudo Rip Models (PR) where the Hubble rate either becomes infinite for large cosmic time or evolves to a de Sitter space. The Little Rip and Pseudo Rip models have been investigated in recent times by many authors. Contreras et al. have obtained some mathematical conditions to link some LR and PR models with some usual cosmological models proposed for regular early universe \cite{Cont2018}. Albarran et al. addressed the quantization of some little siblings of the Big Rip (LSBR) abrupt event with a  phantom fluid \cite{Albaran2018, Albaran2017}. Brevik and co workers have investigated different aspects of viscous LR and PR models \cite{Brevik2011, Brevik2013, Brevik2012, Brevik2012a}. Thermodynamics of LR cosmology in the framework of $f(R,T)$ gravity has been studied by Houndjo et al. \cite{Houndjo2014}.

In the present work, we are interested to investigate some phantom models without any finite time future singularity in the framework of an extended theory of gravity. For this purpose we employ a simple extended gravity theory on an anisotropic universe.  The paper is organised as follow: in Section II, the basic formalism of $f(R,T)$ gravity and the field equations for LRS Bianchi type $I$ space time have been derived. In Section III, the physical parameters such as the energy conditions and the equation of state (EoS) parameter both for anisotropic and isotropic cases are presented. Four different phantom models leading not to finite future Big Rip are investigated in Section IV. Wormhole solutions to these non singular phantom models are obtained in Section-V. Also the possibility of occurrence of Big Trip in wormholes are discussed. At the end the conclusion and summary are presented in Section-VI.

\section{Basic Formalism}
The action for a geometrically modified extended theory with a matter-geometry coupling can be written as 
\begin{equation} \label{eq:1}
S=\int d^4x\sqrt{-g}\left[\frac{1}{16\pi} f(R,T)+ \mathcal{L}_m \right],
\end{equation}
where $\mathcal{L}_m$ is the matter Lagrangian. $f(R,T)$ is an arbitrary function of the Ricci scalar $R$ and the trace $T$ of the energy-momentum tensor. The action is modified geometrically in the sense that if the functional $f(R,T)$ equals to $R$, the action reduces to that of GR. Here the natural unit system  is used where $G=c=1$; $G$ and $c$ are respectively the Newtonian gravitational constant and speed of light in vacuum.

For a minimal matter-geometry coupling within the action, we can split $f(R,T)$ into two distinct functions $f_1(R)$ and $f_2(T)$ so that $f(R,T)=f_1(R)+f_2(T)$. The action for such minimal coupling becomes
\begin{equation} \label{eq:2}
S=\int d^4x\sqrt{-g}\left[\frac{1}{16\pi} \left(f_1(R)+f_2(T)\right)+ \mathcal{L}_m \right].
\end{equation}

Variation of this action with respect to the metric $g_{\mu\nu}$ provides the modified field equation

\begin{equation} \label{eq:3}
R_{\mu\nu}-\frac{1}{2}f^{-1}_{1,R} (R)f_1(R)g_{\mu\nu}=f^{-1}_{1,R}(R)\left[\left(\nabla_{\mu} \nabla_{\nu}-g_{\mu\nu}\Box\right)f_{1,R}(R)+\left[8\pi +f_{2,T}(T)\right]T_{\mu\nu}+\left[f_{2,T}(T)p+\frac{1}{2}f_2(T)\right]g_{\mu\nu}\right].
\end{equation}

In the above, we have assumed that $\mathcal{L}_m=-p$ where $p$ is the pressure of the cosmic fluid and used the shorthand notations:
\begin{equation}\label{eq:4}
f_{1,R} (R)\equiv \frac{\partial f_1(R)}{\partial R},~~~~~~~~~ f_{2,T} (T)\equiv \frac{\partial f_2(T)}{\partial T}, ~~~~~~~~~ f^{-1}_{1,R} (R) \equiv \frac{1}{f_{1,R} (R)}.
\end{equation}
The energy-momentum tensor $T_{\mu\nu}$ is related to the matter Lagrangian as
\begin{equation}\label{eq:5}
T_{\mu\nu}=-\frac{2}{\sqrt{-g}}\frac{\delta\left(\sqrt{-g}\mathcal{L}_m\right)}{\delta g^{\mu\nu}}.
\end{equation}

In order to develop an extended gravity theory from the above  field equation \eqref{eq:3}, we may consider a simple choice $f_1(R)=R$ which provides  GR like field equations
\begin{equation}\label{eq:6}
G_{\mu\nu}= \left[8\pi +f_{2,T}(T)\right]T_{\mu\nu}+\left[f_{2,T}(T)p+\frac{1}{2}f_2(T)\right]g_{\mu\nu},
\end{equation}
which can also be written as
\begin{equation}\label{eq:7}
G_{\mu\nu}= \kappa_{T}\left[T_{\mu\nu}+ T^{int}_{\mu\nu}\right].
\end{equation}
Here, $G_{\mu\nu}= R_{\mu\nu}-\frac{1}{2}Rg_{\mu\nu}$ is the usual Einstein tensor and $\kappa_{T}= 8\pi +f_{2,T}(T)$ is the redefined Einstein constant. $f_{2,T}(T)$ and consequently $\kappa_{T}$  become constants for a linear  functional $f_2(T)$. However, $\kappa_{T}$ evolves with time and dynamically mediates the coupling between the geometry and matter for any non linear choices of the functional $f_{2}(T)$. In \eqref{eq:7}, we have 
\begin{equation}\label{eq:8}
T^{int}_{\mu\nu}=\left[ \frac{f_{2,T}(T)p+\frac{1}{2}f_2(T)}{8\pi +f_{2,T}(T)}\right]g_{\mu\nu},
\end{equation}
which is the effective energy-momentum tensor generated due to the geometrical modification through a minimal coupling with matter. If we drop the $T$ dependent part of the functional $f(R,T)$, this interaction contribution to the energy-momentum tensor will vanish. In other words, a minimal coupling of matter with geometry in the action will behave like an additional matter field which may be responsible to provide an acceleration. This interesting coupling of matter and curvature is motivated from quantum effects and leads to a non vanishing divergence of the energy-momentum tensor $T_{\mu\nu}$. Suitable choice of the functional $f_2(T)$ may lead to viable cosmological model in conformity with recent observations concerning late time cosmic acceleration.

In the present work, we are interested to investigate some little rip cosmologies in the extended gravity theory and for this purpose we consider a linear functional

\begin{equation}\label{eq:9}
\frac{1}{2}f_2(T)=\beta T +\Lambda_0,
\end{equation}
so that
\begin{eqnarray}
\kappa_T &=& 8\pi+2\beta,\label{eq:110}\\
T^{int}_{\mu\nu} &=& \frac{g_{\mu\nu}}{\kappa_T}\left[\left(2p+T\right)\beta+\Lambda_0\right].\label{eq:11}
\end{eqnarray}
The simple minimal coupling of the functional $f(R,T)=f(R)+f(T)$ with linear functions of $f(R)$ and $f(T)$ have been widely used in literature \cite{Harko2011, Mishra18a, Mishra18b, Mishra18c, Shabani2014, Shamir2015, Das2017,  Moraes2017, Deb2018, Yousaf2018, Sharif2019}. Moreover, Ordines and Calson  have  recently constrained this coupling parameter $\beta$ from the observational data on earth's atmosphere \cite{Ordines2019}. One interesting aspect of the present model is that, GR can be easily recovered  for $\beta=0$ and the responsibility of late time cosmic acceleration is shouldered by the constant $\Lambda_0$. In view of this, we may associate $\Lambda_0$ with the usual cosmological constant in GR.

 We chose the anisotropic metric
 
 \begin{equation}\label{eq:12}
ds^2 = dt^2 - A^2dx^2- B^2(dy^2+dz^2),
\end{equation}
where $A=A(t)$ and $B=B(t)$ are the directional scale factors that govern the rates of expansion along different spatial directions. For isotropic case, we assume $A(t)=B(t)=a(t)$, so that the metric reduces to that of the flat FRW model. For the purpose of the present study, we consider the universe to be filled with a cloud of one dimensional cosmic strings with string tension density $\xi$ aligned along the $x$-axis. The energy-momentum tensor for such a fluid is given by
\begin{equation}\label{eq:13}
T_{\mu\nu}=(p+\rho)u_{\mu}u_{\nu} - pg_{\mu\nu}-\xi x_{\mu}x_{\nu},
\end{equation}
with
\begin{equation}\label{eq:14}
u^{\mu}u_{\mu}=-x^{\mu}x_{\mu}=1
\end{equation}
and 
\begin{equation}\label{eq:15}
u^{\mu}x_{\mu}=0.
\end{equation}
 
Here, $\rho$ represents the energy density and is composed of the particle energy density $\rho_p$ and the string tension density $\xi$ so that $\rho=\rho_p+\xi$. It is worth to mention here that, for an isotropic universe with $A(t)=B(t)$, the string tension density $\xi$ vanishes.\\

The field equations in the extended gravity theory can be written as
\begin{eqnarray}
6(k+2)\dot{H}+27H^2 &=& (k+2)^2\left[-\alpha(p-\xi) +\rho \beta+\Lambda_0\right], \label{eq:16}\\
3(k^2+3k+2)\dot{H}+9(k^2+k+1)H^2 &=& (k+2)^2\left[ -\alpha p +(\rho+\xi)\beta+\Lambda_0\right],\label{eq:17}\\
9(2k+1)H^2 &=& (k+2)^2\left[\alpha \rho -\left(p-\xi\right)\beta +\Lambda_0\right].\label{eq:18}
\end{eqnarray}

Here $\alpha=8\pi+3\beta$ and we denote the ordinary time derivatives as overhead dots. The parameter $k$ is a measure of the anisotropic behaviour of the model. We recover an isotropic model for $k=1$, otherwise the model retains its anisotropic nature with asymmetric expansion along the longitudinal and transverse directions. The Hubble parameter $H$ is given by $H=\frac{\dot{a}}{a}=\frac{1}{3}\left(\frac{\dot{A}}{A}+2\frac{\dot{B}}{B}\right)$, $a$ is the scale factor of the universe. Other relevant geometrical quantities include 

\begin{eqnarray}
\text{Expansion scalar:}~~~ \theta &=& u_{;l}^l=\left(\frac{\dot{A}}{A}+2\frac{\dot{B}}{B}\right),\label{eq:19}\\
\text{Deceleration parameter:}~~~ q &=&  -1+\frac{d}{dt}\left(\frac{1}{H}\right),\label{eq:20}\\
\text{Jerk parameter:}~~~  j &=&  \frac{\dddot{a}}{aH^3}=\frac{\ddot{H}}{H^3}-(2+3q).\label{eq:21}
\end{eqnarray}
\section{Physical parameters}
In this section, we wish to derive the expressions of the dynamical physical parameters for both the anisotropic and isotropic models. The anisotropic universe considered in the present work is more general than the FRW model for any values of the parameter $k$. It can be easily reduced to a flat FRW model for $k=1$.
\subsection{Anisotropic case}
 Initally, without putting any restriction on the parameter $k$, the physical properties of the model such as pressure, energy density and string tension density can be obtained from the field equations \eqref{eq:16}-\eqref{eq:18} in terms of the Hubble parameter, the anisotropic parameter $k$ and the coupling parameter as

\begin{eqnarray} 
p &=& -\frac{1}{\kappa_T(\kappa_T+2\beta)}\left[\phi_1(k,\beta)\dot{H}+\phi_2(k,\beta)H^2-\kappa_T\Lambda_0\right],\\ \label{eq:22}
\rho &=& \frac{1}{\kappa_T(\kappa_T+2\beta)}\left[\phi_3(k,\beta)\dot{H}+\phi_4(k,\beta)H^2-\kappa_T\Lambda_0\right],\\\label{eq:23}
\xi &=& \frac{1}{\kappa_T} \left[\phi_5(k)\left(\dot{H}+3H^2\right)\right],\label{eq:24}
\end{eqnarray}

The equation of state parameter (EoS), $\omega=\frac{p}{\rho}$ can be obtained from the above expressions as
\begin{equation}\label{eq:25}
\omega=-1+\frac{\left[\phi_3(k,\beta)-\phi_1(k,\beta)\right]\dot{H}+\left[\phi_4(k,\beta)-\phi_2(k,\beta)\right]H^2}{\phi_3(k,\beta)\dot{H}+\phi_4(k,\beta) H^2-\kappa_T\Lambda_0}.
\end{equation}

Here 
\begin{eqnarray}
\phi_1(k,\beta) &=& \left[8(k+1)\pi+2(2k+1)\beta\right]\chi(k),\label{eq:26}\\
\phi_2(k,\beta) &=& \left[8(k^2+k+1)\pi+(5k^2+3k+1)\beta\right]\chi^2(k),\label{eq:27}\\
\phi_3(k,\beta) &=& -2\beta\chi(k),\label{eq:28}\\
\phi_4(k,\beta) &=& \left[8(2k+1)\pi+(8k+1)\beta\right]\chi^2(k),\label{eq:29}\\
\phi_5(k) &=& (1-k)\chi(k),\label{eq:30}
\end{eqnarray}
where $\chi(k)=\frac{3}{k+2}$. From the expressions of the pressure and energy density we can have
\begin{equation}\label{eq:31}
\rho+p=\frac{1}{\kappa_T(\kappa_T+2\beta)}\left[\left(\phi_3(k,\beta)-\phi_1(k,\beta)\right)\dot{H}+\left(\phi_4(k,\beta)-\phi_2(k,\beta)\right)H^2\right].
\end{equation}

One can note that for $\beta=-2\pi$, we have $\phi_1(k,\beta)=\phi_3(k,\beta)$ and $\phi_2(k,\beta)=\phi_4(k,\beta)$. Consequently in the limit $\beta\rightarrow -2\pi$, a $\Lambda$CDM model is recovered with $p=-\rho$ and $\omega=-1$. In phantom models, one can have $\dot{H} >0, t>0$ and hence the weak energy condition $\rho+p\geq 0; ~ \rho\geq 0$ is not satisfied. It is evident from equation \eqref{eq:31}, a violation of weak energy condition depends on the choice of the parameter $k$ and $\beta$. In view of the recent observations on cosmic anisotropy, there should be almost null departure from the assumption of cosmological principle. This necessitates us to chose a $k$ which envisages an anisotropic universe but with a little departure from isotropy.  Since in the present work we are interested in phantom models, we compel the other parameters appearing in the expression \eqref{eq:31} so that the models violate the weak energy condition.

In the GR limit with $\beta \rightarrow 0$, we have $\phi_3(k,0)-\phi_1(k,0)=-8(k+1)\pi\chi(k)$ and $\phi_4(k,0)-\phi_2(k,0)=8\pi k(1-k)\chi^2(k)$ and consequently, the EoS parameter becomes
\begin{equation}\label{eq:32}
\omega=-1+\frac{(k+1)\chi^{-1}(k)\dot{H}+k(k-1)H^2}{\chi^{-2}(k)\Lambda_0-(2k+1)H^2}.
\end{equation}

In the absence of a cosmological constant it becomes
\begin{equation}\label{eq:33}
\omega=-1-\frac{(k+1)\chi^{-1}(k)\dot{H}+k(k-1)H^2}{(2k+1)H^2}.
\end{equation}

The EoS parameter depends on the anisotropic parameter $k$, the coupling constant $\beta$ besides its dependence on the parameters appearing in the Hubble parameter. It is interesting to note that, the EoS parameter becomes a non evolving parameter in the absence of a cosmological constant for similar time dependence of $\dot{H}$ and $H^2$. However, in the presence of a cosmological constant, it evolves with time.

\subsection{Isotropic case}
We recover the isotropic model for $k=1$ which on substitution in the equations \eqref{eq:26}-\eqref{eq:30} gives
\begin{eqnarray}
\phi_1(\beta) &=& 2(8\pi+\beta),\nonumber\\
\phi_2(\beta)&=& \phi_4(\beta) = 3(8\pi+3\beta),\nonumber\\
\phi_3(\beta) &=& -2\beta,\nonumber\\
\phi_5 &=& 0.\label{eq:34}
\end{eqnarray}

The EoS parameter for the isotropic case becomes
\begin{equation}\label{eq:35}
\omega=-1+8(2\pi+\beta)\frac{\dot{H}}{2\beta\dot{H}-3(8\pi+3\beta)H^2+\kappa_T\Lambda_0}
\end{equation}

In the limit $\beta \rightarrow 0$ and $\Lambda_0 \rightarrow 0$, the EoS parameter reduces to that of the FRW model
\begin{equation}\label{eq:36}
\omega=-1-\frac{2}{3}\frac{\dot{H}}{H^2}.
\end{equation}

The weak energy condition in the isotropic case becomes
\begin{equation}\label{eq:37}
\rho+p= -\frac{2\dot{H}}{(\kappa_T+2\beta)}.
\end{equation}

Since in phantom models, we have $\dot{H}>0$ in positive time frame, violation of the weak energy condition is clearly visible in this model for positive coupling constant $\beta$.
\section{Rip cosmologies}
We wish to investigate some rip cosmologies in the extended gravity theory both for anisotropic and isotropic universes. For this purpose we restrict ourselves to specific Little Rip(LR) models. Little Rip cosmologies are very interesting where the Hubble rate tends to infinity at an infinite time. An interesting fact in these models is that the EoS parameter asymptotically and sufficiently rapidly reaches to $-1$ \cite{Framp2011}. Singularities occur in these models but at an  infinite future. In other words, there is no effective singularity. 
\subsection{Little Rip}
The Hubble parameter for the LR model can be taken as \cite{Framp2011, Framp2012}
\begin{equation}\label{eq:39}
H=H_0e^{\lambda t},~~~~~~~~~~ H_0>0,~~\lambda >0 
\end{equation}
so that the scale factor is expressed as a double exponential expression
\begin{equation}
a = a_0~ exp\left[\frac{H_0}{\lambda}\left(e^{\lambda t}-e^{\lambda t_0}\right)\right].
\end{equation}
Here $a_0$ is the scale factor at the present epoch $t_0$. In this case, the Hubble rate increases exponentially with time and thereby produces strong inertial force. With the growth of cosmic time, the inertial force increases and any bound system tends to rip at an infinitely large time. In this kind of model, rip occurs but not at a finite time, a phenomenon dubbed as Little Rip.

The deceleration parameter and the jerk parameter for the LR scale factor are expressed as

\begin{eqnarray}
q &=& -1-\frac{\lambda}{H_0}e^{-\lambda t},\label{eq:40}\\
j &=& 1+\frac{3\lambda}{H_0}e^{-\lambda t}+\left(\frac{\lambda}{H_0}\right)^2e^{-2\lambda t}.\label{eq:41}
\end{eqnarray}
The deceleration parameter and the jerk parameter asymptotically approach to $-1$ and $1$ respectively. At the present epoch, the deceleration parameter $q$ becomes $q_0=-1-\frac{\lambda}{H_0}e^{-\lambda t_0}$ which implies that $q_0 <-1$. In $\Lambda$CDM model, the jerk parameter at the present epoch is predicted to $j_0=1$. The jerk parameter for the LR model, at the present epoch, has a value given by $j_0=1+\frac{3\lambda}{H_0}e^{-\lambda t_0}+\left(\frac{\lambda}{H_0}\right)^2e^{-2\lambda t_0}$. This value is greater than that predicted from $\Lambda$CDM model.   

Since $\dot{H}=\lambda H >0$, substitution of equation \eqref{eq:39} into equation \eqref{eq:25}, yields the EoS parameter for the LR model as
\begin{equation}\label{eq:42}
\omega_{LR}=-1+\frac{\left[\phi_3(k,\beta)-\phi_1(k,\beta)\right]\lambda H^{-1}+\left[\phi_4(k,\beta)-\phi_2(k,\beta)\right]}{\phi_3(k,\beta)\lambda H^{-1}+\phi_4(k,\beta) -\kappa_T\Lambda_0H^{-2}},
\end{equation}
which can be explicitly expressed as 
\begin{equation}\label{eq:43}
\omega_{LR}=-1+\frac{\left[\phi_3(k,\beta)-\phi_1(k,\beta)\right]\frac{\lambda}{H_0}e^{-\lambda t}+\left[\phi_4(k,\beta)-\phi_2(k,\beta)\right]}{\phi_3(k,\beta)\frac{\lambda}{H_0}e^{-\lambda t}+\phi_4(k,\beta) -\kappa_T\frac{\Lambda_0}{H_0^2}e^{-2\lambda t}}.
\end{equation}

The evolution of the EoS parameter in the LR model depends on the anisotropic parameter $k$, the coupling constant $\beta$, the parameters of the scale factors $\lambda$ and $H_0$.
At an initial epoch, $t\rightarrow 0$, we have
\begin{equation}\label{eq:44}
\omega_{LR}= -1+\frac{\left[\phi_3(k,\beta)-\phi_1(k,\beta)\right]\frac{\lambda}{H_0}+\left[\phi_4(k,\beta)-\phi_2(k,\beta)\right]}{\phi_3(k,\beta)\frac{\lambda}{H_0}+\phi_4(k,\beta) -\kappa_T\frac{\Lambda_0}{H_0^2}},
\end{equation} 
and at a late phase ($t\rightarrow \infty$)
\begin{equation}\label{eq:45}
\omega_{LR}(t\rightarrow \infty)=-\frac{\phi_2(k,\beta)}{\phi_4(k,\beta)}.
\end{equation}
It is obvious that, the model evolves in a phantom phase with $\omega_{LR}<-1$ at an initial epoch to $\omega_{LR} \rightarrow -1$ at late phase thereby holding the LR scenario. However, the asymptotic value of the EoS depends on the anisotropic parameter $k$ and the coupling constant $\beta$.

For an isotropic case, the evolutionary behaviour of the EoS parameter in the LR model is given by
\begin{equation}\label{eq:46}
\omega^{iso}_{LR}=-1+\frac{8(2\pi+\beta)\frac{\lambda}{H_0}e^{-\lambda t}}{2\beta\frac{\lambda}{H_0}e^{-\lambda t}-3(8\pi+3\beta)+\kappa_T\frac{\Lambda_0}{H_0^2}e^{-2\lambda t}},
\end{equation}
which asymptotically approaches to $-1$ as $t\rightarrow \infty$. In the limit of GR with $\beta\rightarrow0$ and $\Lambda_0\simeq 0$, we have 
\begin{equation}\label{eq:47}
\omega^{iso(GR)}_{LR}=-1-\frac{2}{3}\frac{\lambda}{H_0}e^{-\lambda t}.
\end{equation}

\subsection{Pseudo Rip}
Another phantom behaviour without singularity at finite time is speculated by a Hubble parametrization \cite{Framp2012}
\begin{equation}\label{eq:48}
H=H_0-H_1e^{-\lambda t},
\end{equation}
$H_0, H_1$ and $\lambda$ are positive constants and $H_0 > H_1$.  Since, in the limit $t\rightarrow +\infty$, the Hubble parameter becomes a constant $H\rightarrow H_0$, this model evolves asymptotically to a de Sitter universe. Such a model corresponds to a Pseudo Rip (PR) model.  The first derivative of the Hubble rate becomes $\dot{H}=\lambda H_1e^{-\lambda t}=\lambda(H_0-H) >0$. The scale factor for this Hubble parameter can be obtained as
\begin{equation}\label{eq:49}
a=a_0~exp\left[H_0(t-t_0)+\frac{H_1}{\lambda}\left(e^{-\lambda t}-e^{-\lambda t_0}\right)\right].
\end{equation}

The deceleration parameter $q$ and the jerk parameter $j$ for the PR model are given by
\begin{eqnarray}
q &=&-1-\frac{\lambda H_1e^{-\lambda t}}{\left(H_0-H_1e^{-\lambda t}\right)^2},\label{eq:50} \\
j &=& 1-\frac{\lambda H_1e^{-\lambda t}\left[\lambda+3(H_0-H_1e^{-\lambda t})\right]}{\left(H_0-H_1e^{-\lambda t}\right)^3}.\label{eq:51}
\end{eqnarray}
While the deceleration parameter at an initial epoch is $q(t\rightarrow 0)=-1-\frac{\lambda H_1}{\left(H_0-H_1\right)^2}$, it approaches $-1$ at late times. The deceleration parameter in general evolves from a higher negative value to $-1$ at late epoch. On the otherhand, the jerk parameter evolves from a low value of $j=1-\frac{\lambda H_1\left[\lambda+3(H_0-H_1)\right]}{\left(H_0-H_1\right)^3}$ to $j=1$ at late times. However, these parameters will have singularities at $t=ln\left(\frac{H_1}{H_0}\right)^{\frac{1}{\lambda}}$.

The EoS parameter for the PR model can be obtained as
\begin{equation}\label{eq:52}
\omega_{PR}=-1+\frac{\left[\phi_3(k,\beta)-\phi_1(k,\beta)\right]\lambda H_1e^{-\lambda t}+\left[\phi_4(k,\beta)-\phi_2(k,\beta)\right]\left(H_0-H_1e^{-\lambda t}\right)^2}{\phi_3(k,\beta)\lambda H_1e^{-\lambda t}+\phi_4(k,\beta) \left(H_0-H_1e^{-\lambda t}\right)^2-\kappa_T\Lambda_0}.
\end{equation}

The evolution of the EoS parameter in the PR model depends on the anisotropic parameter $k$, the coupling constant $\beta$, the parameters of the scale factors $\lambda$ and $H_0$.
At an initial epoch, $t\rightarrow 0$, we have
\begin{equation}\label{eq:53}
\omega_{PR}(t\rightarrow 0)= -1+\frac{\left[\phi_3(k,\beta)-\phi_1(k,\beta)\right]\lambda H_1+\left[\phi_4(k,\beta)-\phi_2(k,\beta)\right]\left(H_0-H_1\right)^2}{\phi_3(k,\beta)\lambda H_1+\phi_4(k,\beta) \left(H_0-H_1\right)^2-\kappa_T\Lambda_0},
\end{equation} 
and at a late phase ($t\rightarrow \infty$)
\begin{equation}\label{eq:54}
\omega_{PR}(t\rightarrow \infty)=-1+\frac{\left[\phi_4(k,\beta)-\phi_2(k,\beta)\right]H_0^2}{\phi_4(k,\beta)H_0^2-\kappa_T\Lambda_0}.
\end{equation}
However, in the absence of a cosmological constant, it reduces to $\omega_{PR}(t\rightarrow \infty)=-\frac{\phi_2(k,\beta)}{\phi_4(k,\beta)}.$
It is obvious that, this pseudo rip model evolves in a phantom phase with $\omega_{LR}<-1$ at an initial epoch to $\omega_{LR} \rightarrow -1$ at late phase. Just like the little rip case, in this model also, the asymptotic value of the EoS depends on the anisotropic parameter $k$ and the coupling constant $\beta$.

In order to understand the evolutionary behaviour of the EoS parameter for the PR model in an isotropic universe, we need to substitute $k=1$ and can obtain in a straightforward way
\begin{equation}\label{eq:55}
\omega^{iso}_{PR}=-1+\frac{8(2\pi+\beta)\lambda H_1e^{-\lambda t}}{2\beta\lambda H_1e^{-\lambda t}-3(8\pi+3\beta)\left(H_0-H_1e^{-\lambda t}\right)^2+\kappa_T\Lambda_0},
\end{equation}
which asymptotically approaches to $-1$ as $t\rightarrow \infty$. In the limit of GR with $\beta\rightarrow0$ and $\Lambda_0\simeq 0$, we have 
\begin{equation}\label{eq:56}
\omega^{iso(GR)}_{PR}=-1-\frac{2}{3}\frac{\lambda H_1e^{-\lambda t}}{\left(H_0-H_1e^{-\lambda t}\right)^2}.
\end{equation}

The phantom evolution of the EoS parameter is obvious. It evolves from $\omega_{PR} <-1$ to an asymptotic value of $-1$. One can note that this model has a $\omega$-singularity at $t=t_{\omega}=ln\left(\frac{H_1}{H_0}\right)^{\frac{1}{\lambda}}$ in the framework of GR.
\subsection{Emergent Little Rip}
We may consider a scale factor describing an emergent solution as considered by Mukherjee et al.\cite{Mukh2006}
\begin{equation}\label{eq:57}
a(t)=a_i\left(\nu+e^{\mu t}\right)^{\gamma},
\end{equation}
where $a_i, \mu, \nu$ and $\gamma$ are positive constants.

The Hubble parameter for this ansatz is given by
\begin{equation}\label{eq:58}
H(t)=\frac{\mu\gamma e^{\mu t}}{\nu+e^{\mu t}}.
\end{equation}
It is obvious that as $t\rightarrow \infty$, we have $a\rightarrow\infty$ and $H\rightarrow \mu\gamma$. This model asymptotically evolves to a de Sitter universe. Also we have 
\begin{equation}\label{eq:59}
\dot{H}=\frac{\mu\gamma e^{\mu t}}{\nu+e^{\mu t}}\left[\mu-\frac{1}{\gamma}\frac{\mu\gamma e^{\mu t}}{\nu+e^{\mu t}}\right]=H\left(\mu-\frac{H}{\gamma}\right)
\end{equation}
which in the limit $t\rightarrow \infty$ approaches to $0$. The value of the parameters $\mu$ and $\gamma$ are chosen in such a manner that, $\dot{H}>0$ for $t>0$.

The deceleration parameter and the jerk parameter for this emergent little rip (ELR) model are obtained as
\begin{eqnarray}
q &=& -1+\frac{1}{\gamma}-\frac{\nu+e^{\mu t}}{\gamma e^{\mu t}},\label{eq:60}\\
j &=& \left(1-\frac{3}{\gamma}+\frac{2}{\gamma^2}\right)+\frac{\mu}{H}+\frac{\mu(\mu-2/\gamma)}{H^2}.\label{eq:61}
\end{eqnarray}
While the deceleration parameter evolves from $q=-1-\frac{\nu}{\gamma}$ to $-1$, the jerk parameter evolves from $j=1+\frac{\nu-2}{\gamma}+\frac{1}{\gamma^2}\left[2+\frac{(\nu+1)^2(\mu-2/\gamma)}{\mu}\right]$ at an initial phase  to $1-\frac{2}{\gamma}+\frac{1}{\gamma^2}\left[2+\frac{(\mu-2/\gamma)}{\mu}\right]$ at late phase of evolution.\\

For the ELR model with the Hubble parameter in equation \eqref{eq:58}, the EoS parameter turns out to be
\begin{equation}\label{eq:62}
\omega_{ELR}=-1+\frac{\left[\phi_3(k,\beta)-\phi_1(k,\beta)\right]\left(\frac{\mu}{H}-\frac{1}{\gamma}\right)+\left[\phi_4(k,\beta)-\phi_2(k,\beta)\right]}{\phi_3(k,\beta)\left(\frac{\mu}{H}-\frac{1}{\gamma}\right)+\phi_4(k,\beta)-\kappa_T\Lambda_0 H^{-2}}.
\end{equation}

This EoS parameter evolves in phantom phase with $\omega_{ELR} <-1$ and asymptotically reduces to 

\begin{equation}\label{eq:63}
\omega_{ELR}(t\rightarrow \infty)=-1+\frac{\left[\phi_4(k,\beta)-\phi_2(k,\beta)\right]}{\phi_4(k,\beta)-\frac{\kappa_T\Lambda_0}{\mu^2\gamma^2}}
\end{equation}
at a late epoch ($t\rightarrow \infty$). In this model also, the EoS parameter reduces to $\omega_{ELR}(t\rightarrow \infty)=-\frac{\phi_2(k,\beta)}{\phi_4(k,\beta)}$ in the absence of a cosmological constant.

For an isotropic universe, we can have
\begin{equation}\label{eq:64}
\omega^{iso}_{ELR}=-1+\frac{8(2\pi+\beta)\left(\frac{\mu}{H}-\frac{1}{\gamma}\right)}{2\beta\left(\frac{\mu}{H}-\frac{1}{\gamma}\right)-3(8\pi+3\beta)+\kappa_T\Lambda_0 H^{-2}},
\end{equation}
which asymptotically approaches to $-1$ as $t\rightarrow \infty$. In the limit of GR with $\beta\rightarrow0$ and $\Lambda_0\simeq 0$, we have 
\begin{equation}\label{eq:65}
\omega^{iso(GR)}_{ELR}=-1-\frac{2}{3}\left(\frac{\mu}{H}-\frac{1}{\gamma}\right).
\end{equation}

\subsection{Bouncing with Little Rip}
Myrzakulov and Sebastini \cite{Myrza2014} have studied a scale factor in exponential form 
\begin{equation}\label{eq:66}
a(t)=a_0e^{(t-t_0)^{2n}},
\end{equation}
where $a_0>0$ is the scale factor at time $t_0$. The exponent $n\neq 0$ decides the bouncing behaviour of the model.

The Hubble parameter for this ansatz is given by
\begin{equation}\label{eq:67}
H(t)= 2n(t-t_0)^{2n-1},
\end{equation}
so that its first derivative becomes $\dot{H}=2n(2n-1)(t-t_0)^{2n-2}$. For $t>0$, the condition $\dot{H}>0$ requires that $n>\frac{1}{2}$. This model provides a little rip at late times when the exponent $n$ assumes positive integral numbers. The model bounces at $t=t_0$ when the bouncing scale factor becomes $a_0$. It is obvious that as $t\rightarrow \infty$, we have $a\rightarrow\infty$ and $H\rightarrow \infty$ for positive integral values of $n$.

The deceleration parameter and the jerk parameter for this bouncing with little rip (BLR) model are obtained as
\begin{eqnarray}
q &=& -1-\frac{2n-1}{2n(t-t_0)^{2n}},\label{eq:68}\\
j &=& 1+\frac{3(2n-1)}{2n(t-t_0)^{2n}}+\frac{(n-1)(2n-1)}{2n^2(t-t_0)^{4n}}.\label{eq:69}
\end{eqnarray}
The deceleration parameter is a negative quantity for $n > \frac{1}{2}$ and evolves to an asymptotic value of $q=-1$. The jerk parameter evolves to $j=1$ at late times.

For the BLR model we can calculate the EoS parameter as
\begin{equation}\label{eq:70}
\omega_{BLR}=-1+\frac{\left[\phi_3(k,\beta)-\phi_1(k,\beta)\right]\left[\frac{2n-1}{2n(t-t_0)^{2n}}\right]+\left[\phi_4(k,\beta)-\phi_2(k,\beta)\right]}{\phi_3(k,\beta)\left[\frac{2n-1}{2n(t-t_0)^{2n}}\right]+\phi_4(k,\beta)-\frac{\kappa_T\Lambda_0}{4n^2(t-t_0)^{4n-2}}}
\end{equation}

which asymptotically reduces to $\omega_{BLR}(t\rightarrow \infty)=-\frac{\phi_2(k,\beta)}{\phi_4(k,\beta)}$.

The EoS parameter for this BLR model in an isotropic universe can be expressed as, 
\begin{equation}\label{eq:71}
\omega^{iso}_{BLR}=-1+\frac{8(2\pi+\beta)\left[\frac{2n-1}{2n(t-t_0)^{2n}}\right]}{2\beta\left[\frac{2n-1}{2n(t-t_0)^{2n}}\right]-3(8\pi+3\beta)+\frac{\kappa_T\Lambda_0}{4n^2(t-t_0)^{4n-2}}},
\end{equation}
\section{Wormhole Solutions and Big Trip}
The phantom energy accretion on to wormhole leads to an increase in the size of the wormhole throat which may eventually engulf the whole universe before the occurrence of any kind of rip. Such a phenomenon is called Big Trip. In this section, we will calculate the wormhole throat radius and its evolution under the phantom energy accretion.

The throat radius $R(t)$ of Morris-Thorne wormhole can be calculated for phantom dark energy models from the evolution equation \citep{Asta2012, Babichev2004}

\begin{equation}\label{eq:73}
\dot{R}= -CR^2(\rho+p).
\end{equation}
Here $C$ is a positive dimensionless constant. In our discussion, we will restrict ourselves only to the isotropic case with vanishing cosmological constant.\\

\textit{Case-I: Little Rip:}
For the LR model in equation \eqref{eq:39}, $\dot{H}=\lambda H_0e^{\lambda t}$, we obtain from equation \eqref{eq:37}
\begin{equation}\label{eq:74}
\rho+p=-\frac{2\lambda H_0e^{\lambda t}}{\kappa_T+2\beta}
\end{equation}

Substituting \eqref{eq:74} into \eqref{eq:73}, we get the wormhole throat radius for the LR case as
\begin{equation}
\frac{1}{R_{LR}(t)}= -\frac{2C}{\kappa_T+2\beta}H_0e^{\lambda t} +k_1,\nonumber
\end{equation}
where $k_1$ is an integration constant. 

At Big Trip, $t=t_B$ and we have $k_1=\frac{2C}{\kappa_T+2\beta}H_0e^{\lambda t_B}$. Hence one obtains  the wormhole throat radius for the LR model as
\begin{equation}\label{eq:75}
R_{LR}(t)=\frac{\kappa_T+2\beta}{2CH_0}\left[e^{\lambda t_B}-e^{\lambda t}\right]^{-1}.
\end{equation}

Assuming the wormhole throat radius at $t=t_0$ to be $R_0$, we may have the Big Trip at an epoch

\begin{equation}\label{eq:76}
t_B=ln\left[e^{\lambda t_0}+\frac{\kappa_T+2\beta}{2CH_0R_0}\right]^{\frac{1}{\lambda}}.
\end{equation}
In the limit, $\beta\rightarrow 0$, the extended gravity theory reduces to GR so that the redefined Einstein constant becomes $\kappa_T= \kappa_T+2\beta= 8\pi$. Therefore in the GR limit, we can have
\begin{equation}\label{eq:77}
t_B=ln\left[e^{\lambda t_0}+\frac{8\pi}{2CH_0R_0}\right]^{\frac{1}{\lambda}}.
\end{equation}
A comparison of \eqref{eq:76} and \eqref{eq:77} shows that, the presence of a positive finite coupling constant in the extended gravity increases the time of occurrence of the Big Trip.\\

\textit{Case-II: Pseudo Rip:}
The throat radius for this case is obtained as
\begin{equation}
R_{PR}(t)=\frac{\kappa_T+2\beta}{2CH_1}\left[e^{-\lambda t}-e^{-\lambda t_B}\right]^{-1}
\end{equation}
and consequently the Big Trip occurs at
\begin{equation}
t_B=ln\left[e^{-\lambda t_0}-\frac{\kappa_T+2\beta}{2CH_1R_0}\right]^{-\frac{1}{\lambda}}.
\end{equation}
It is interesting to note that, Big Trip occurs for the wormholes if their radius at $t=t_0$ satisfy the condition
\begin{equation}
R_0 >\frac{\kappa_T+2\beta}{2CH_1}e^{\lambda t_0}.
\end{equation}

\textit{Case-III: Emergent Little Rip:}
In this case, the Hubble parameter is given by equation \eqref{eq:58} and we get the wormhole throat radius as
\begin{equation}
R_{ELR}(t)=\frac{\kappa_T+2\beta}{2C\mu\nu\gamma}\left[\frac{1}{\nu+e^{\mu t}}-\frac{1}{\nu+e^{\mu t_B}}\right]^{-1}.
\end{equation}

For this emergent little rip model, the Big Trip occurs at
\begin{equation}
t_B=ln\left[\left(\frac{1}{\nu+e^{\mu t_0}}-\frac{\kappa_T+2\beta}{2C\mu\nu\gamma R_0}\right)^{-1}-\nu\right]^{\frac{1}{\mu}}.
\end{equation}

In this case, the Big Trip will occur if the wormhole simultaneously satisfies the conditions
\begin{equation}
R_0 > \frac{(\kappa_T+2\beta)(\nu+e^{\mu t_0})}{2C\mu\nu\gamma}
\end{equation}
and 
\begin{equation}
R_0<\frac{\nu(\kappa_T+2\beta)}{2C\mu\nu\gamma\left[\nu-(\nu+e^{\mu t_0})\right]}.
\end{equation}

\textit{Case-IV: Bouncing with Little Rip:}
A Bouncing model with little rip behaviour may be obtained for the Hubble parametrization given in equation \eqref{eq:67}.

The wormhole throat radius for this model becomes
\begin{equation}
R_{BLR}(t)=\frac{\kappa_T+2\beta}{4Cn}\left[\left(t_B-t_0\right)^{2n-1}-\left(t-t_0\right)^{2n-1}\right]^{-1},
\end{equation}
and the Big Trip occurs at
\begin{equation}
t_B=t_0+\left[\left(t^{\prime}-t_0\right)^{2n-1}+\frac{\kappa_T+2\beta}{4CnR^{\prime}}\right]^{\frac{1}{2n-1}}.
\end{equation}
Here $t_0$ is the bouncing epoch and $t^{\prime}$ is the time corresponding to the wormhole radius $R^{\prime}$.
\section{Conclusion}In the present work, we have discussed some phantom models without Big Rip singularity at finite future in the frame work of an extended theory of gravity. The extended theory of gravity is derived from an action where the usual Ricci Scalar is replaced by a coupled function which is linear in $R$ and $T$. The presence of the trace of the energy momentum tensor in the geometry side of the Einstein-Hilbert action provides an acceleration. We have constructed some phantom models with $\omega$ evolving from $\omega <-1$ to an asymptotic value of $-1$ at an infinitely large time. Such models are called Little Rip models where the Hubble rate approaches to large values at infinitely large time. In some cases, the Hubble parameter evolves to a de Sitter space. In these models, singularity does not occur effectively. 

We have investigated four different models of Little Rip or Pseudo Rip both for anisotropic and isotropic universe and obtained the dynamical evolution of the EoS parameter. Also we have investigated the violation of energy conditions in these models. The model parameters are found to affect substantially the dynamical behaviour of the EoS parameter. 

Wormhole solutions are obtained for the four Little Rip and Pseudo Rip models. It is certain that, it is possible to obtain wormhole solutions in phantom models where Big Rip can be avoided. The time of Big Trip for all the rip models considered in the work have been calculated. It is demonstrated that, a modified gravity theory such as the present one delays the time of occurrence of Big Trip in wormholes than that in GR. In Pseudo Rip models, Big Trip can occur with certain limiting conditions for the wormhole throat. We have obtained those conditions for our models.
Phantom models are always interesting and can be confronted with observations. With a lot of observational data coming in recent times that favour a phantom world, we hope, our theoretical models within a simple extended gravity theory may be quite useful. 

\section*{Acknowledgement}
BM and SKT thank IUCAA, Pune (India) for hospitality and support during an academic visit where a part of this work is accomplished. 

\end{document}